\documentclass[twocolumn]{article}
\usepackage[margin=2cm]{geometry}
\usepackage{titlesec}
\titleformat{\section}{\normalfont\bfseries\filcenter\MakeUppercase}{\thesection.}{1em}{}
\renewcommand{\thesection}{\Roman{section}}
\titleformat{\subsection}{\normalfont}{\itshape\thesubsection.}{1em}{\itshape\/}
\renewcommand{\thesubsection}{\Alph{subsection}}
\newcommand{\figfontsize}{\fontsize{8pt}{9.5pt}\selectfont} % Il faudra \figfontsize dans chaque environnement de figure où on veut l'appliquer (on peut l'automatiser avec floatrow, mais ça génère plein de soucis et de conflits, et, en plus, il est probable qu'il y a ait des figure où l'on ne veut pas s'en servir).
\newcommand{\lmrfont}{\fontfamily{lmr}\selectfont}
\usepackage{caption} % Pour gérer les cartouches de figures et tables
\DeclareCaptionLabelFormat{smallscformat}{{\lmrfont\textsc{#1 #2}}}
\DeclareCaptionFormat{smallformat}{{\figfontsize#1#2#3}}
\usepackage{graphicx}
\usepackage{amsmath, amsfonts}
\usepackage{mathrsfs}
\graphicspath{{figures/},{../note/figures}}
\usepackage{color}
\definecolor{pyC0}{RGB}{31,119,180} % La couleur 'C0' de python (bleu)",
\definecolor{pyC1}{RGB}{255,127,14} % La couleur 'C1' de python (orange)
\definecolor{pyC2}{RGB}{44,160,44}
\usepackage{authblk}
\usepackage[style=ieee,backend=biber,datamodel=software]{biblatex}
\usepackage{software-biblatex}
\bibliography{biblio_cylinders.bib}
\usepackage{anyfontsize} % lmodern not sufficient, I had to use anyfontsize
\usepackage{booktabs}
\usepackage{multirow}
\usepackage{stackengine}

\begin{document}
    \title{\Large{\textbf{Electromagnetic Scattering by a Finite Metallic Circular Cylinders Set}}}
    \date{}

    \author[1,*]{Matthieu Elineau}
    \author[1]{Lucille Kuhler}
    \author[1]{Alexandre Chabory}
    \affil[1]{ENAC lab, Université de Toulouse, Toulouse, France}
    \affil[*]{e-mail: matthieu.elineau@proton.me}

    \renewcommand\Authfont{\fontsize{11}{13.2}\selectfont}
    \renewcommand\Affilfont{\fontsize{9}{10.8}\selectfont}
    
    \renewcommand{\abstractname}{\vspace{-3\baselineskip}}
    
    \twocolumn[
	\begin{@twocolumnfalse}
		\maketitle
		\begin{abstract}
			The problem of electromagnetic scattering by cylinders is an old problem that has been studied in many configurations. The present publication provides a theoretical study on a not yet investigated general case: the set of finite metallic circular cylinders. A model which takes into account both the finiteness of the cylinders and their electromagnetic coupling is provided. The total field is written in a two dimensional problem in terms of cylindrical harmonics and is used to define current densities which are integrated in a three dimensional problem. The finiteness is taken into account assuming current densities that are identical from those of the two dimensional problem. Coupling effects are naturally taken into account \textit{via} the matrix formulation of the boundary condition that binds together the cylindrical harmonic coefficients. The proposed closed-form is valid for great cylinder lengths and any cylinder radii. Numerical experiments are also provided in various configurations in order to evaluate the accuracy of the model. The model computational times happens to be 5 orders of magnitude shorter than a full-wave reference simulation, without significant loss of accuracy.\vspace{1\baselineskip}
		\end{abstract}
	\end{@twocolumnfalse}
	]

    \section{Introduction}
        Many structures involved in an electromagnetic scattering problem can be approximated with canonical shapes in order to ease the modelling of the situation. Along with spherical or planar structures, cylindrical\footnote{Throughout the article, for the sake of brevity, the word ``cylinder'' is used to designate the phrase ``circular cross section cylinder''. } bodies are frequently found in electromagnetic environments. Situations can range from stubs in a waveguide to wind turbines in an outdoor propagation environment. For immense structures such as the last mentioned, efficient models are mandatory before considering simulations.

        Analytical solutions which are nowadays numerically very frugal has been found many years ago for the simple case of the infinite cylinder. A solution exists for the metallic cylinder since at least the beginning of the 20th century with Rayleigh's work~\cite{Rayleigh1918}. Since then, the problem has been solved for the general case of oblique incidence for dielectric cylinders~\cite{Wait1955}.

        The problem has also been extended to finite cylinders, \textit{i.e.} cylinders generated by the translation of a circle along a finite segment. For small angles of incidence~\cite{Lind1966} and for various radii with integral equation techniques~\cite{Shepherd1983}. The general case of the finite length dielectric and arbitrary oriented cylinder has been solved~\cite{Seker1988} by calculation of the current densities for the infinite cylinder followed by the calculation of radiation integrals on finite surfaces. Extended boundary condition method has also allowed to describe the field scattered by a finite cylinder by the means of a spherical vector wavefunctions expansion~\cite{Ruppin1990}. A global study encompasses all possible regimes for finite circular metallic cylinders by specifying all the closed form solutions when possible~\cite{Brill1995}.
        \begin{figure}
            \centering
            \includegraphics{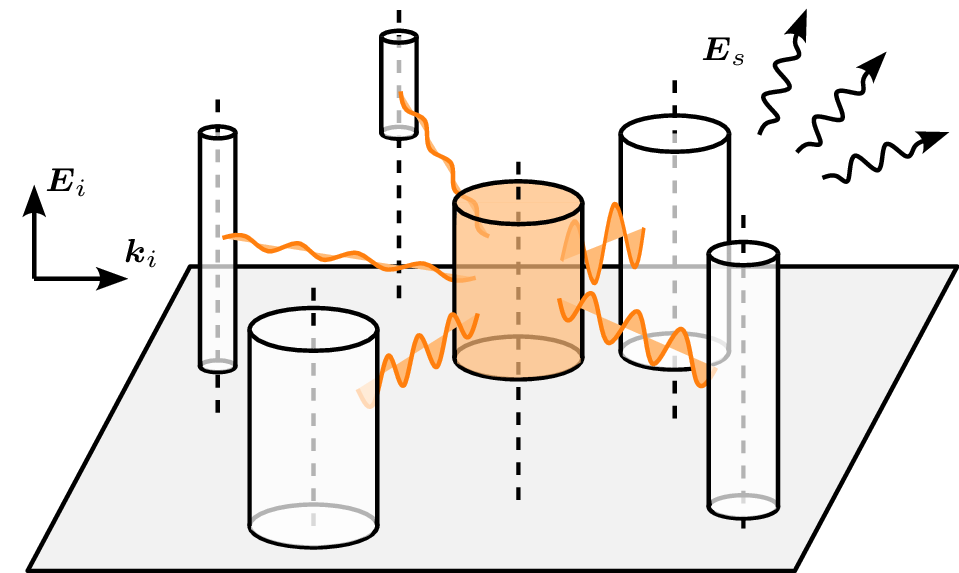}
            \caption{Illustration of the scattering by a set of cylinders. They are finite and of arbitrary radius, height and position. The scattered field is the result of the interaction between the incident wave and the coupled cylinders.}\label{fig_illustration_problem}
        \end{figure}

        Few decades after the first studies on individual cylinders, a strong interest grown toward the problem of sets of cylinders. The use of the cylindrical wave expansion in combination with addition theorems through a matrix formulation then became quite common~\cite{Ragheb1985,Elsherbeni1992,Felbacq1994}. This last technique has been extended to the description of arbitrary cross section cylinders with the equal volume technique. Periodic arrangements of cylinders has been extended to multilayered structures~\cite{Yasumoto2004} as well as lacunar structures~\cite{Watanabe2012}. A very general work described the scattering from a set of arbitrary positioned dielectric cylinders, under oblique incidence~\cite{Henin2007}. More recent studies explored some applications for ferromagnetic materials~\cite{Kumar2015} or magnetized plasma~\cite{Ivoninsky2017}.

        The objective of this article is to give a model for the combination of both the finite cylinder case and the set of infinite cylinders case. The solution is provided for an arbitrary arrangement of co-oriented finite metallic cylinders of arbitrary radius and arbitrary but large lengths for a parallel to the cylinders electric polarisation, as illustrated in Figure~\ref{fig_illustration_problem}. The model combines both the coupling between cylinders \textit{via} the summation with Graf's theorem of the cylindrical harmonics and the finiteness of the cylinders \textit{via} the calculation of radiation integrals on finite surfaces.
        
        The outline of the article is the following. In the first part of the model development, in Section~\ref{sec_model_2d}, infinite cylinders are considered in order to define the total field in terms of cylindrical harmonics. The boundary condition of the problem is written in a matrix formulation, which allows to write a linear system that leads to the cylindrical harmonic coefficients values when solved. In the second part of the model development, in Section~\ref{sec_model_3d}, a current density is calculated from the total field expression in the two dimensional configuration. This current density is used in the calculation of radiation integrals over the finite surface of finite cylinders, \textit{i.e.} in a three dimensional configuration. Currents are therefore assumed to be identical in both situations, which is considered true for great cylinder lengths. The integration gives as many scattered fields as cylinders, which can be summed in order to finally give the total scattered field. The model is implemented and compared with full wave simulations in Section~\ref{sec_validation}.

    \section{Two dimensional total field}\label{sec_model_2d}
        Starting the model development in a two dimensional (\textit{i.e.} invariant along a given space direction) configuration allows to write the fields in terms of cylindrical harmonics. With such a decomposition, coupling interactions between the cylinders of a given set are taken into account. After defining the problem configuration, this section details the calculation of the total field and the associated cylindrical harmonic coefficients.
        
        \subsection{Two dimensional problem configuration}
\begin{figure}[t]
                \centering
                \includegraphics{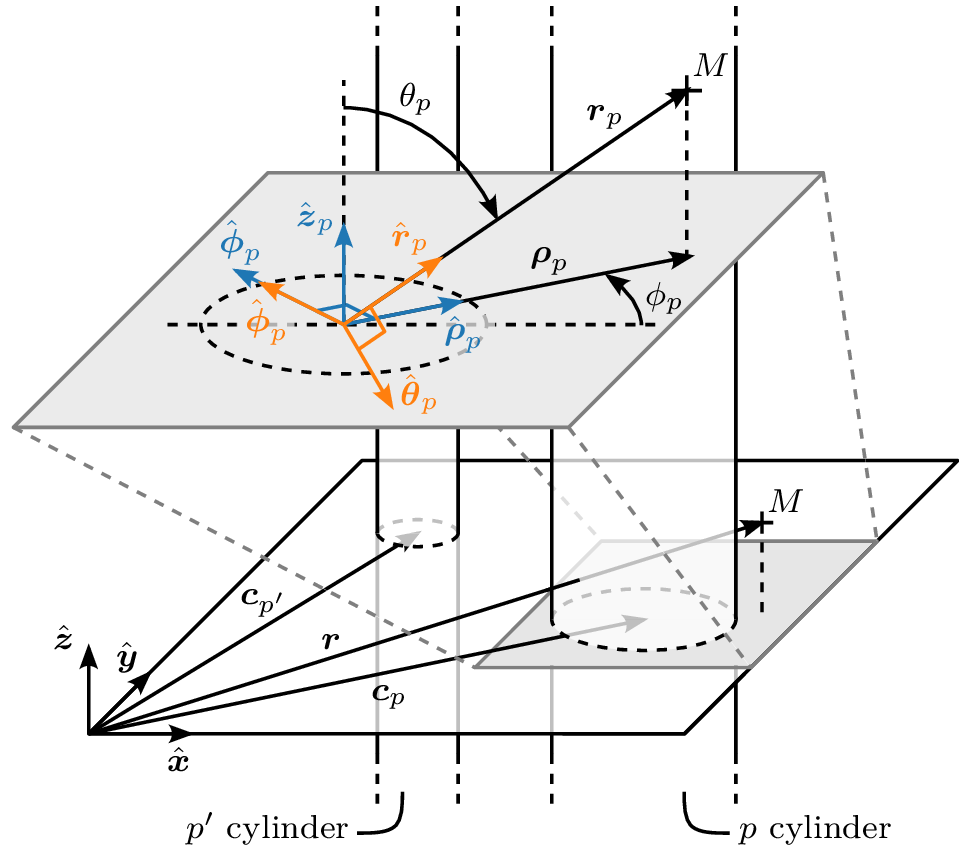}
                \caption{The two dimensional configuration which consists of a set of $\hat{\boldsymbol{z}}$ oriented infinite cylinders. Each cylinder is associated with both cylindrical local basis and spherical local basis which are represented in the zoomed rectangle.}\label{fig_infinite_conf}
            \end{figure}

            The problem is chosen to be $\hat{\boldsymbol{z}}$ invariant, as described in Figure~\ref{fig_infinite_conf}. The configuration consists of a set of $P$ $\hat{\boldsymbol{z}}$-oriented cylinders illuminated by a $\hat{\boldsymbol{z}}$ polarized plane wave propagating along $\hat{\boldsymbol{x}}$. A $\boldsymbol{c}_p$ vector with $p \in \{1, 2, \dots, P\}$ designates the center of a cylinder $p$ as well as the origin of a cylindrical $\{\hat{\boldsymbol{\rho}}_p, \hat{\boldsymbol{\phi}}_p, \hat{\boldsymbol{z}}_p\}$ local basis and a spherical $\{\hat{\boldsymbol{r}}_p, \hat{\boldsymbol{\theta}}_p, \hat{\boldsymbol{\phi}}_p\}$ local basis. All the $\boldsymbol{c}_p$ vectors are contained in a $z=0$ plane which serves as a reference plane for the three dimensional case later. Any $M$ point can either be designated in the local cylindrical and spherical bases or in the global cartesian basis by a vector $\boldsymbol{r}$. It is assumed that every calculations are performed in the context free space propagation.
        
        \subsection{Fields in local bases in the two dimensional problem}
            The total field $\boldsymbol{E}$ is the Maxwell's equations solution, the incident field $\boldsymbol{E}_i$ is the field without the scatterers and the scattered field $\boldsymbol{E}_s$ is the difference $\boldsymbol{E} - \boldsymbol{E}_i$. Both incident and scattered fields can be decomposed in cylindrical harmonics in each cylindrical basis. Each point where they are evaluated is then located by the $(\rho_p, \phi_p, z_p)$ coordinate triplet. 

            \paragraph{Incident field} For a $\hat{\boldsymbol{z}}$ polarized plane wave of wavenumber $k$ propagating along $\hat{\boldsymbol{x}}$, the cylindrical harmonic expansion~(5--101) in~\cite{Harrington2001} in the local basis of the cylinder $p$ is
            \begin{equation}\label{equ_inc_field}
                E_{i, z_p}(\rho_p, \phi_p) = \sum_n s_n^{(p)}J_n(k\rho_p)e^{jn\phi_p}
            \end{equation}
            where $J_n$ is the $n$ order Bessel function of the first kind and $s_n^{(p)}$ are the coefficients of the expansion which are, for the present case,
            \begin{equation}
                s^{(p)}_n = E_{i0}e^{-jkx_{cp}}j^{-n}.
            \end{equation}
            $E_{i0}$ is the amplitude of the incident wave and $x_{cp}$ is the $\hat{\boldsymbol{x}}$ component of the $\boldsymbol{c}_p$ vector. The $e^{+j\omega t}$ convention is assumed throughout the article.
            
            \paragraph{Scattered field} The scattered field is a sum of $P$ components, each of them being associated with a cylinder $p$. At any point outside cylinder $p$ the scattered field can be represented by the out-going harmonics associated with $p$. Summing the contribution of all cylinders, this leads to an expression for the scattered field outside the cylinders, which is given by (see equation~(5--106) from~\cite{Harrington2001} for example)
            \begin{equation}\label{equ_E_zs_notok}
                E_{s, z_p} = \sum_{p=1}^{P}\sum_{n=-\infty}^{\infty}a_n^{(p)}H_n^{(2)}(k\rho_p)e^{jn\phi_p}.
            \end{equation}
            where $H_n^{(2)}$ is the $n$-th order Hankel function of the second kind. This expression is nevertheless not suitable for applying the boundary condition at the surface of each cylinder because it uses a different cylindrical coordinate system for each cylinder. It is therefore needed to expand the scattered field using only one coordinate system. This can be done by first considering the out-going harmonics of a local basis $p'$ as in-going harmonics in the local basis $p$. Using Bessel functions of the first kind to model in-going harmonics, this gives
            \begin{equation}\label{equ_scat_field}
                \begin{split}
                    E_{s, z_p}(\rho_p, \phi_p) =& \sum_n a_n^{(p)}H_n^{(2)}(k\rho_p)e^{jn\phi_p} \\
                    &+ \sum_{p'\neq p} \sum_n b_n^{(p, p')} J_n(k\rho_p)e^{jn\phi_p}
                \end{split}
            \end{equation}
            where $b_n^{(p, p')}$ are the amplitudes of the in-going harmonics yielded by the scattering of cylinder $p'$ on cylinder $p$. The coefficient $b_n^{(p, p')}$ contains the same information as an $a_n^{(p')}$: the scattering of cylinder $p'$. They are linked by the means of the Graf's addition theorem~(10.23.7) in~\cite{Olver2010} which gives, for our case,
            \begin{equation}
                \begin{split}
                    &H_{n'}^{(2)}(k\rho_{p'})e^{jn'\phi_{p'}} = \\ 
                    &\quad \sum_{n} H_{n'-n}^{(2)}\left(kR^{(p, p')}\right) J_n(k\rho_p) e^{jn\phi_p}e^{j(n'-n)\Phi^{(p, p')}},
                \end{split}
            \end{equation}
            where $\Phi^{(p,p')} = \angle(\boldsymbol{c}_p - \boldsymbol{c}_{p'}, \hat{\boldsymbol{x}})$ and $R^{(p, p')} = \lVert\boldsymbol{c}_p - \boldsymbol{c}_{p'}\rVert$ are the cylindrical coordinates of the cylinder $p$ center designated by $\boldsymbol{c}_p$, in the $p'$ cylinder centered frame whose origin is at $\boldsymbol{c}_{p'}$. Writing
            \begin{equation}\label{equ_m_elements}
                M_{n, n'}^{(p, p')} = H_{n'-n}^{(2)}\left(kR^{(p, p')}\right)e^{j(n'-n)\Phi^{(p,p')}}
            \end{equation}
            allows to identify the product $M_{n, n'}^{(p, p')}a_n^{(p')}$ in the scattered field~(\ref{equ_scat_field}) which becomes
            \begin{equation}\label{equ_scat_field_a_only}
                \begin{split}
                    E_{s, z_p}(\rho_p, \phi_p) =& \sum_n a_n^{(p)}H_n^{(2)}(k\rho_p)e^{jn\phi_p} \\
                    &+ \sum_{p'\neq p} \sum_n M_{n, n'}^{(p, p')}a_n^{(p')} J_n(k\rho_p)e^{jn\phi_p}.
                \end{split}
            \end{equation}
            This is an expression of the scattered field with the $a_n^{(p)}$ coefficients as only unknowns.
            
            \paragraph{Total field} The total field is simply obtained from the summation of the incident field~(\ref{equ_inc_field}) and the scattered field~(\ref{equ_scat_field}). This summation gives
            \begin{equation}\label{equ_Ez}
                \begin{split}
                    E_{z_p}(\rho_p, \phi_p) = \sum_n \Biggr[\Biggr(s_n^{(p)} + \sum_{p'\neq p} M_{n, n'}^{(p, p')}a_n^{(p')}\Biggr)J_n(k\rho_p) \\
                    +\ a_n^{(p)}H_n^{(2)}(k\rho_p)\Biggr]e^{jn\phi_p}.
                \end{split}
            \end{equation}

        \subsection{Boundary condition}
            The total field can be expressed in terms of $a_n^{(p)}$ only by using the boundary condition on the surface of a metallic cylinder ($E_{z_p} = 0$). Nullifying the left hand side of the total field formulation~(\ref{equ_Ez}) and considering that every set of functions under summation terms over $n$ are orthogonal on $[0, 2\pi]$ this gives, on cylinder $p$,
            \begin{equation}\label{equ_bc_orgn}
                a_n^{(p)} + \frac{J_n(kR_p)}{H_n^{(2)}(kR_p)}\sum_{p'\neq p}^{} M_{n, n'}^{(p, p')}a_n^{(p')} = -s_n^{(p)}\frac{J_n(kR_p)}{H_n^{(2)}(kR_p)}.
            \end{equation}
            This boundary condition can alternatively be written
            \begin{equation}\label{equ_bc}
                s_n^{(p)}+\sum_{p'\neq p}^{}M_{n, n'}^{(p, p')}a_n^{(p')} = -a_n^{(p)}\frac{H_n^{(2)}(kR_p)}{J_n(kR_p)}
            \end{equation}
            and inserted into the total field formulation~(\ref{equ_Ez}) to give
            \begin{equation}\label{equ_Ez_only_a}
                \begin{split}
                    E_{z_p} = \sideset{}{'}\sum_n a_n^{(p)}\Biggr[H_n^{(2)}(k\rho_p) - J_n(k\rho_p)\frac{H_n^{(2)}(kR_p)}{J_n(kR_p)}\Biggr]e^{jn\phi_p}.
                \end{split}
            \end{equation}
            This expression is the total field in the cylinder $p$ local basis and only implies $a_n^{(p)}$ coefficients. It can be expressed in any cylinder $p$ centered frame and the $p'$ indices designate another cylinder of the set. One can notice that the denominator of the ratio in this equation can be evaluated to zero since the Bessel first kind functions have zeros. In such very special cases, the boundary condition~(\ref{equ_bc_orgn}) gives $a_n^{(p)} = 0$ and this is clear from that same boundary condition that the product $M_{n, n'}^{(p, p')}a_n^{(p')}$, which is also $b_n^{(p, p')}$, is not constrained and can therefore take any value. In such a case, the coefficients are associated to resonant modes in the cylinder which are not to be taken into account for the later calculation of a scattered field. We therefore introduced the $'$ notation in order to specify that the summation is to be performed on $n \in \mathbb{N} \setminus n | J_n(kR_p) = 0$. One can still safely rely on the initial formulation~(\ref{equ_Ez}) but this formulation is heavier for the later developments.
        
        \subsection{Matrix formulation}\label{subsec_matrix_form}
            In practice, only a limited number of cylindrical function orders $n$ is needed. They are usually restricted to the first integer greater than two times the number of wavelengths in a cylinder section perimeter, which is $N^{(p)} = \lceil 2kR_p\rceil$ for each cylinder $p$. Column vectors of coefficients are introduced for the unknown $a_n^{(p)}$ and the source term $s_n^{(p)}$, considering a finite number of possible integer values for $n$ with $n\in[-N^{(p)}, N^{(p)}]$ such that
            \begin{equation}
                \boldsymbol{a}^{(p)} = \begin{bmatrix} 
                    a_{-N^{(p)}}^{(p)} \\
                    \vdots \\
                    a_{N^{(p)}}^{(p)}
                    \end{bmatrix},\quad
                \boldsymbol{s}^{(p)} = \begin{bmatrix} 
                    s_{-N^{(p)}}^{(p)} \\
                    \vdots \\
                    s_{-N^{(p)}}^{(p)} 
                    \end{bmatrix},
            \end{equation}
            The introduction of the diagonal matrix
            \begin{equation}
                \boldsymbol{\Lambda}^{(p)} = 
                        \begin{bmatrix}
                            \frac{J_{-N^{(p)}}(kR_p)}{H_{-N^{(p)}}^{(2)}(kR_p)} &  & \\
                            & \ddots & \\
                            & & \frac{J_{N^{(p)}}(kR_p)}{H_{N^{(p)}}^{(2)}(kR_p)}
                        \end{bmatrix}
            \end{equation}
            allows to write the boundary condition~(\ref{equ_bc_orgn}) in matrix form with $a_n^{(p)}$ as the only unknown such that
            \begin{equation}\label{equ_system_p}
                \boldsymbol{a}^{(p)} + \boldsymbol{\Lambda}^{(p)} \sum_{p'=1}^{P} \boldsymbol{M}^{(p, p')}\boldsymbol{a}^{(p')} = -\boldsymbol{\Lambda}^{(p)}\boldsymbol{s}^{(p)}.
            \end{equation}
            The linear system~(\ref{equ_system_p}) holds for the cylinder $p$. $P$ systems can be written and gathered in the total linear system
            \begin{equation}\label{equ_linear_system}
                \boldsymbol{M}\boldsymbol{a} = \boldsymbol{\nu}
            \end{equation}
            where the unknown and sources are
            \begin{equation}
                \boldsymbol{a} = \begin{bmatrix}
                    \boldsymbol{a}^{(1)} \\
                    \vdots \\
                    \boldsymbol{a}^{(P)}
                \end{bmatrix}, \quad
                \boldsymbol{\nu} = -\begin{bmatrix}
                    \boldsymbol{s}^{(1)}\boldsymbol{\Lambda}^{(1)}\\
                    \vdots \\
                    \boldsymbol{s}^{(P)}\boldsymbol{\Lambda}^{(P)}
                \end{bmatrix}
            \end{equation}
            and the bloc matrix is
            \begin{equation}\label{equ_m}
                \boldsymbol{M} = \begin{bmatrix}
                    \boldsymbol{I} & \boldsymbol{\Lambda}^{(1)}\boldsymbol{M}^{(1, 2)} & \cdots & \boldsymbol{\Lambda}^{(1)}\boldsymbol{M}^{(1, P)} \\
                    \boldsymbol{\Lambda}^{(2)}\boldsymbol{M}^{(2, 1)} & \boldsymbol{I} & & \\
                    \vdots & & \ddots & \\
                    \boldsymbol{\Lambda}^{(P)}\boldsymbol{M}^{(P, 1)} & & & \boldsymbol{I}
                \end{bmatrix}.
            \end{equation}
            The unknown $\boldsymbol{a}$ of the linear system~(\ref{equ_linear_system}) depends only on the sources. The solution of this system then leads to the $a_n^{(p)}$ coefficients which in turn gives complete knowledge of the total field~(\ref{equ_Ez_only_a}).

    \section{Three dimensional scattered field}\label{sec_model_3d}
        In this section, the previously introduced two dimensional total field expression~(\ref{equ_Ez_only_a}) is converted into a current density at the surface of each cylinder $p$ and used for the calculation of a radiation integral. The integration is performed over the finite surfaces of the cylinders, providing a three dimensional solution to the problem.
        \subsection{Two dimensional current density}
            At the surface of a $\hat{\boldsymbol{z}}$ oriented cylinder $p$ of radius $R_p$, where a normal unitary vector $\hat{\boldsymbol{n}}_p$ is defined, the current density $\boldsymbol{J}^{(p)}$ is an electric current density which is classically obtained from Maxwell equations, considering the PEC boundary condition and a $\hat{\boldsymbol{z}}$ polarized electric field such that
            \begin{equation}\label{equ_jz_interm}
                J_{z_p}^{(p)} = (\hat{\boldsymbol{n}}_p \times \boldsymbol{H})\cdot \hat{\boldsymbol{z}}_p = \left. \frac{1}{j\omega\mu_0}\frac{\partial E_{z_p}}{\partial\rho_p}\right|_{\rho_p= R_p}
            \end{equation}
            is the only $\boldsymbol{J}^{(p)}$ non null component. This current density notation $J_{z_p}$ is not to be mistaken with the Bessel functions notation $J_n$. The insertion of the total field expression~(\ref{equ_Ez_only_a}) in the current density expression~(\ref{equ_jz_interm}) along with the use of the recurrence relations of the cylindrical functions~(10.6.1) in~\cite{Olver2010} gives
            \begin{equation}\label{equ_jz_article}
                \begin{split}
                    &J_{z_p}^{(p)}(\phi_p) = \frac{k}{2j\omega\mu_0}\\ 
                    &\quad \times \sum_{n}a_n^{(p)}H_n^{(2)}(kR_p)\left[ \mathscr{H}_n(kR_p) - \mathscr{J}_n(kR_p)\right] e^{jn\phi_p}.
                \end{split}
            \end{equation}
            This current density is a sum of cylindrical functions in which the contributions from the Hankel and Bessel functions are separated into two compact quantities that are
            \begin{equation}\label{equ_Hcal}
                \mathscr{H}_n(kR_p) = \frac{H_{n-1}^{(2)}(kR_p) - H_{n+1}^{(2)}(kR_p)}{H_n^{(2)}(kR_p)}
            \end{equation}
            and
            \begin{equation}\label{equ_Jcal}
                \mathscr{J}_n(kR_p) = \frac{J_{n-1}(kR_p) - J_{n+1}(kR_p)}{J_n(kR_p)}.
            \end{equation}
            At this point, the problem is still two dimensional so $\phi_p$ is the only space dependency of $J_{z_p}^{(p)}$.

        \subsection{Three dimensional scattered field in a local basis}
            \begin{figure}
                \centering
                \includegraphics{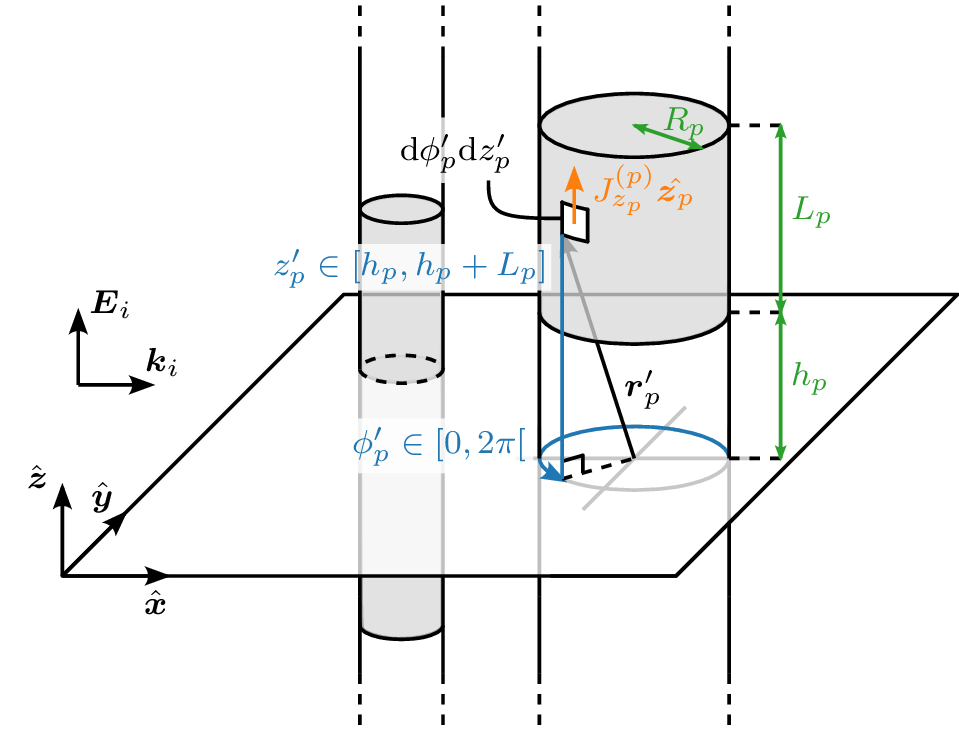}
                \caption{A set of finite cylinders illuminated by a plane wave propagating along $\hat{\boldsymbol{x}}$. This figure is the three dimension extension of Figure~\ref{fig_infinite_conf}. The current density calculated in the 2D configuration is used to calculate the radiation integral over the finite surface of a cylinder.}\label{fig_finite_conf}
            \end{figure}
            In order to obtain the three dimensional scattered field for one cylinder $p$, the final step is to integrate the previously calculated current density $J_{z_p}^{(p)}$ over the surface of this cylinder $p$. We use the classical far field formulation of the Stratton-Chu radiation integrals
            \begin{equation}
                \begin{split}
                    &\boldsymbol{E}_s^{(p)}(r_p, \theta_p, \phi_p) = j\omega\mu_0\frac{e^{-jkr_p}}{4\pi r_p} \hat{\boldsymbol{r}}_p \times \hat{\boldsymbol{r}}_p \\
                    &\times \int_{z_p' = h_p}^{z_p' = h_p+L_p}\int_{\phi_p' = 0}^{\phi_p' = 2\pi} J_{z_p}^{(p)}(\phi_p')\hat{\boldsymbol{z}}_p e^{jk\hat{\boldsymbol{r}}_p\cdot\boldsymbol{r}_p'}R_p\mathrm{d}z_p'\mathrm{d}\phi_p'.
                \end{split}
            \end{equation}
            The integration surface is illustrated in Figure~\ref{fig_finite_conf}. This scattered field is only $\hat{\boldsymbol{\theta}}_p$ oriented so we have ${\boldsymbol{E}_s^{(p)} = E_{s, \theta_p}^{(p)} \hat{\boldsymbol{\theta}}_p}$. After explicitation of $e^{jk\hat{\boldsymbol{r}}_p\cdot\boldsymbol{r}_p'}$ and insertion of the current expression~(\ref{equ_jz_article}) we obtain
            \begin{equation}\label{equ_interm_es}
                \begin{split}
                    &E_{s, \theta_p}^{(p)} = \frac{kR_p}{8\pi} \frac{e^{-jkr_p}}{r_p}\\
                    &\times\sum_{n}\Big\{a_n^{(p)}H_n^{(2)}(kR_p)\left[\mathscr{H}_n(kR_p) - \mathscr{J}_n(kR_p)\right]\\
                    &\times \iint e^{jn\phi_p'} e^{jk[R_p\sin(\theta_p)\cos(\phi_p' - \phi_p) + z_p'\cos(\theta_p)]}\mathrm{d}z_p'\mathrm{d}\phi_p'\Big\}.
                \end{split}
            \end{equation}
            The two integrals are along $z_p'$ and $\phi_p'$ and can be calculated independently. The calculation steps involved are described in Appendix~\ref{supp_radiation_integral} and they lead to the final expression of the scattered field for one cylinder which is
            \begin{equation}\label{equ_e_theta_p}
                E_{s, \theta_p}^{(p)}(r_p, \theta_p, \phi_p) = \mathscr{E}^{(p)}(r_p, \theta_p)\sum_{n}\mathscr{S}_n^{(p)}(\theta_p, \phi_p)
            \end{equation}
            where the amplitude attenuation driven by the cylinder length and the summation term gathering the cylindrical functions are separated into two quantities that are
            \begin{equation}
                \mathscr{E}^{(p)}(r_p, \theta_p) = \frac{k R_p L_p}{4} \frac{\sin(Z_p)}{Z_p}\frac{e^{-jkr_p}}{r_p}e^{jk\cos(\theta_p)\big(h_p+\frac{L_p}{2}\big)}
            \end{equation}
            with $Z_p = \frac{k\cos(\theta_p)L_p}{2}$ and
            \begin{equation}\label{equ_summation_term}
                \begin{split}
                    \mathscr{S}_n^{(p)}(\theta_p, \phi_p) &= a_n^{(p)}H_n^{(2)}(kR_p)\left[\mathscr{H}_n(kR_p) - \mathscr{J}_n(kR_p)\right]\\
                    &\quad\times j^n J_n\big(kR_p\sin(\theta_p)\big)e^{jn\phi_p}.
                \end{split}
            \end{equation}
        
        \subsection{Scattered field for the set of cylinders}
            The $E_{s, \theta_p}^{(p)}(r_p, \theta_p, \phi_p)$ component of the scattered field is used to write the $\boldsymbol{E}_s^{(p)}(\boldsymbol{r}_p)$ scattered field in the local cartesian basis. Any given $M$ point being localized by ${\boldsymbol{r}_p= \boldsymbol{r} - \boldsymbol{c}_p}$ (Figure~\ref{fig_infinite_conf}), the field scattered by the set of cylinders in the global cartesian basis is then
            \begin{equation}\label{equ_total_field}
                \boldsymbol{E}_s(\boldsymbol{r}) = \sum_{p}^{}\boldsymbol{E}_s^{(p)} (\boldsymbol{r}-\boldsymbol{c}_p)
            \end{equation}
            With this expression, once the problem parameters are set ($k$, cylinders dimensions and positions), the scattered field is known everywhere outside the cylinders if the $a_n^{(p)}$ coefficients are known.
    
    \section{Numerical experiments}\label{sec_validation}
        \begin{figure*}
            \centering
            \includegraphics{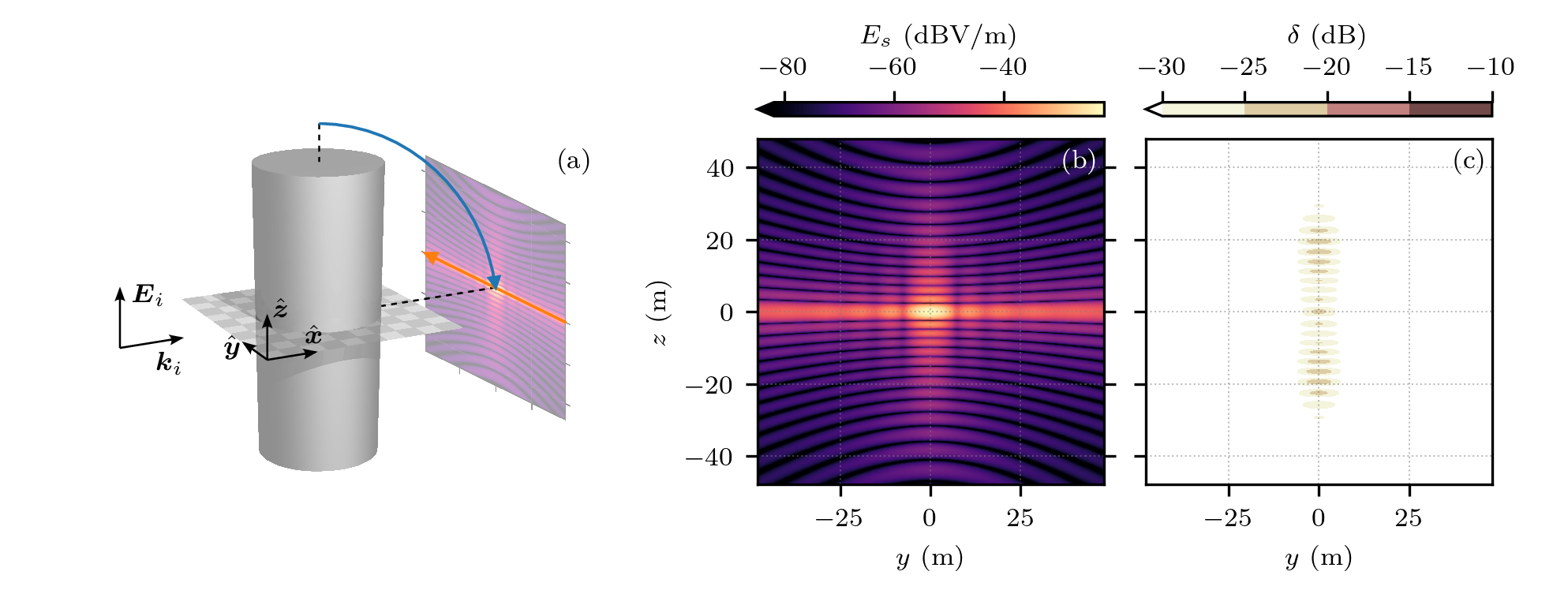}
            \vspace{-3mm}
            \includegraphics{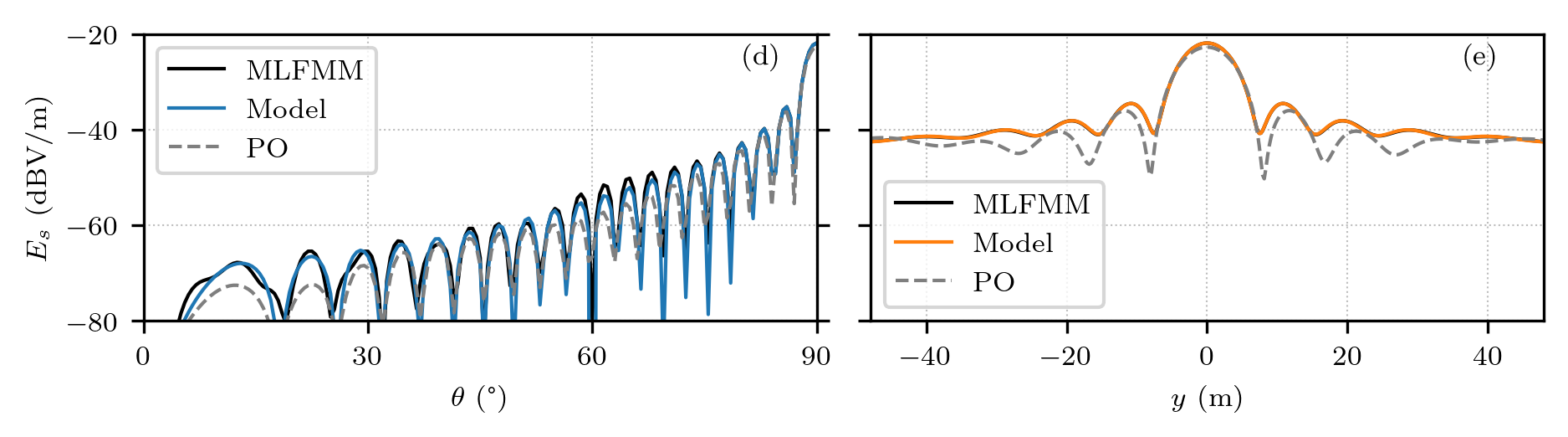}
            \caption{Scattering by a thick cylinder.\ (a) Illumination scene, (b) model, $x = 2d_\mathrm{far}$, (c) error, $x = 2d_\mathrm{far}$, (d) $r = 2d_\mathrm{far}$, $\phi = 0^\circ$ and (e) $z = 0$, $x = 2d_\mathrm{far}$. Blue (\textcolor{pyC0}{---}) and orange (\textcolor{pyC1}{---}) arrows of (a) are the x-axes of (d) and (e) subplots respectively, which are associated with the same model curve color.}\label{fig_thick_cylinder}
        \end{figure*}
        \begin{figure*}
            \centering
            \includegraphics{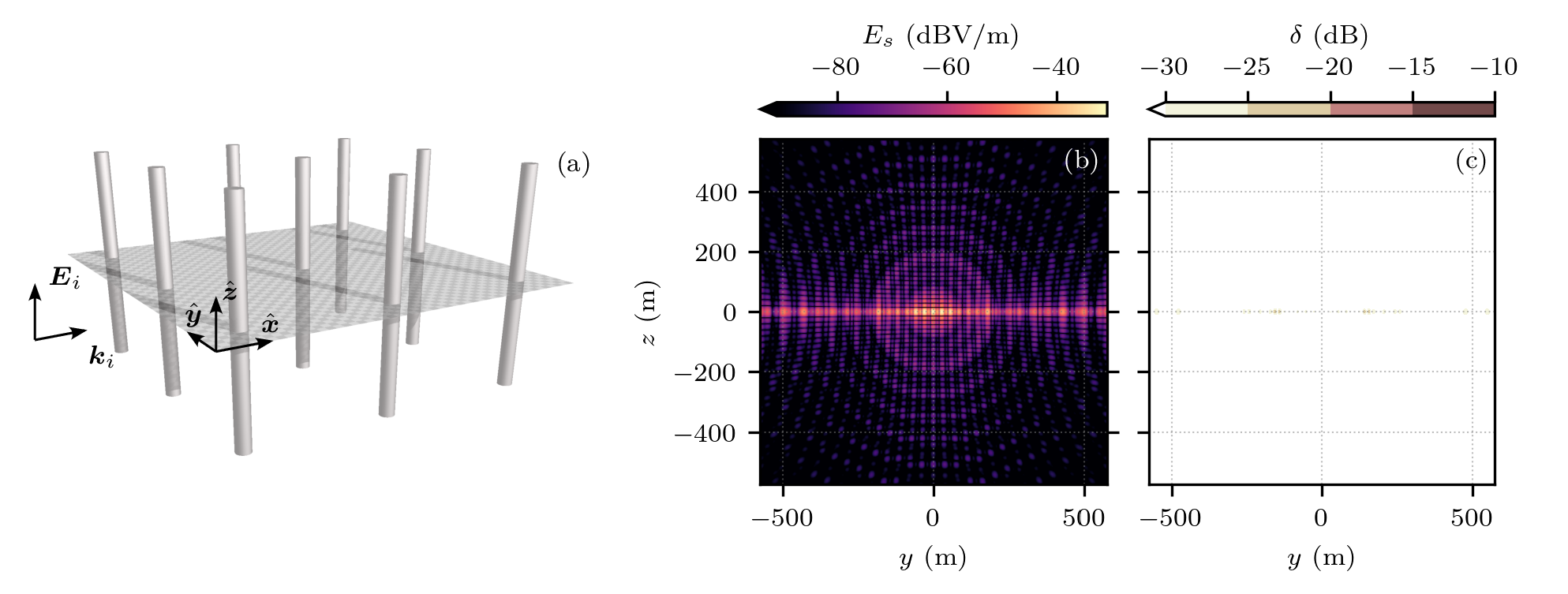}
            \includegraphics{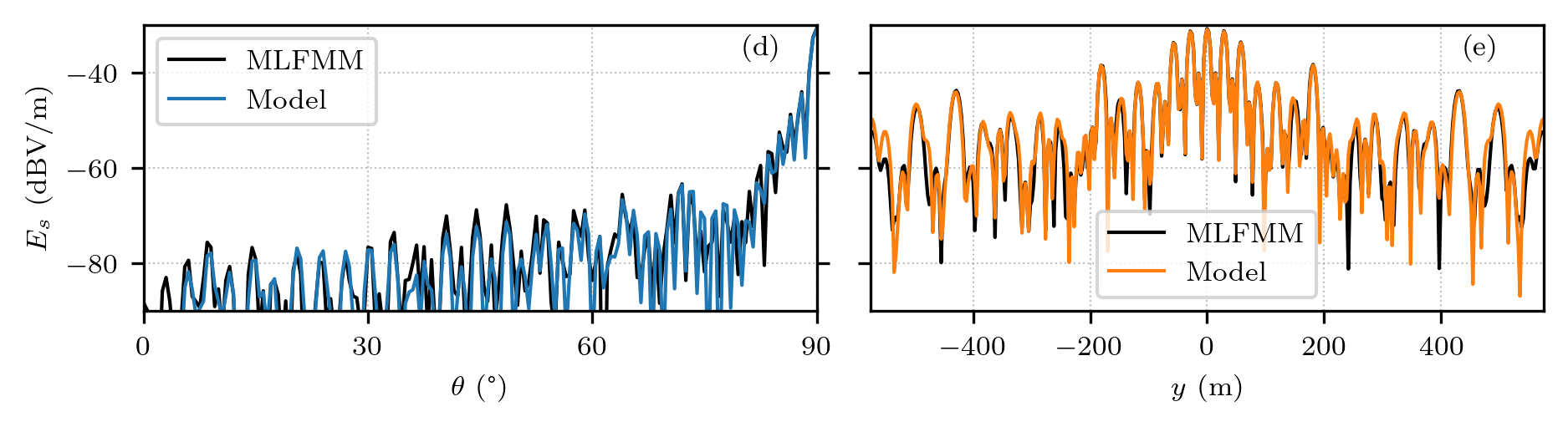}
            \caption{Scattering by a set of 9 identical cylinders.\ (a) Illumination scene, (b) model, $x = 2d_\mathrm{far}$, (c) error, $x = 2d_\mathrm{far}$, (d) $r = 2d_\mathrm{far}$, $\phi = 0^\circ$ and (e) $z = 0$, $x = 2d_\mathrm{far}$.}\label{fig_9_cylinders}
        \end{figure*}
        \begin{figure*}
            \centering
            \includegraphics{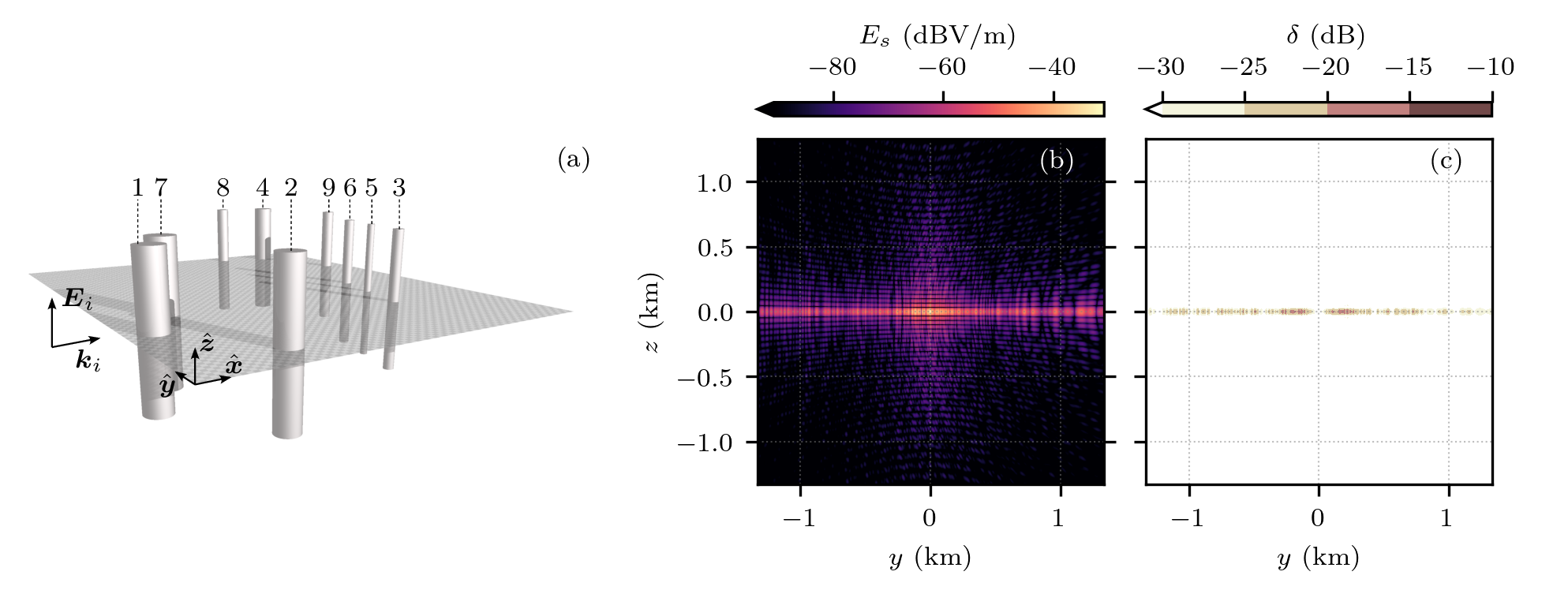}
            \includegraphics{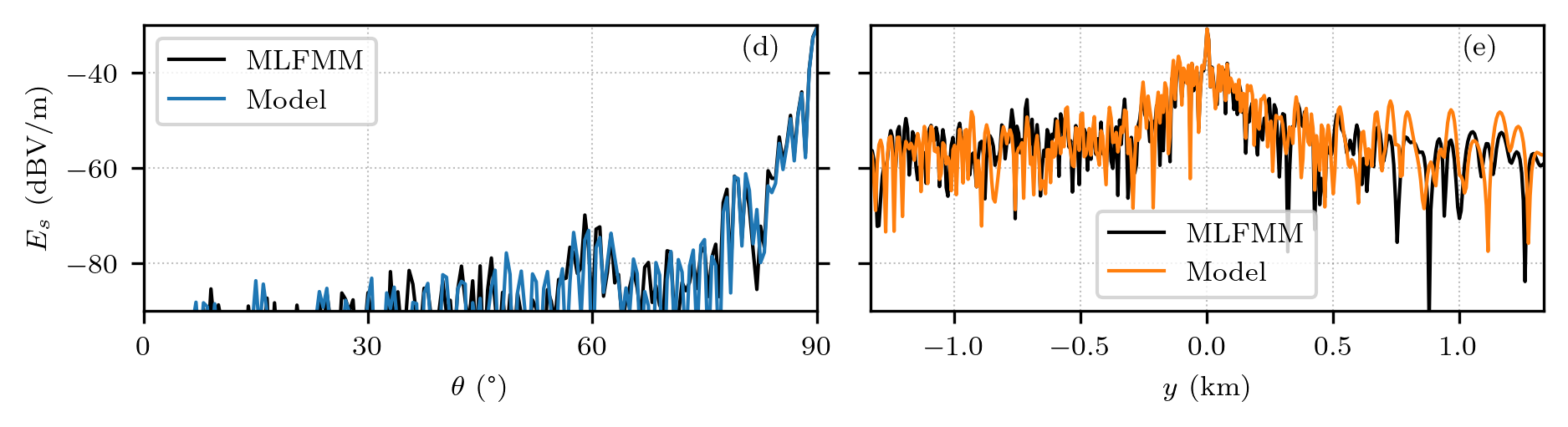}
            \caption{Scattering by a set of 9 randomly placed cylinders of random radii. The index $p$ of each cylinder is specified above the cylinders in the scene.\ (a) Illumination scene, (b) model, $x = 2d_\mathrm{far}$, (c) error, $x = 2d_\mathrm{far}$, (d) $r = 2d_\mathrm{far}$, $\phi = 0^\circ$ and (e) $z = 0$, $x = 2d_\mathrm{far}$.}\label{fig_9_cylinders_randc_randR}
        \end{figure*}
        This section provides comparisons between the proposed model and full wave simulations with a Multi-Level Fast Multipole Method (MLFMM) solver implemented in Feko software in order to evaluate the model accuracy. For our model, the 2 main computation steps are solving the linear system~(\ref{equ_total_field}) so as to obtain the current coefficients, and summing the scattered fields~(\ref{equ_linear_system}). 
        % \textbf{These computations are implemented in an open source python code~\cite{Douvenot2022} code de Rémi pour test, changer.} 
        Although every configuration parameters are expressed in terms of $\lambda$, a frequency $f = 10$~GHz is chosen throughout the entire numerical experiment. Fields are computed beyond the far-field distance $d_\mathrm{far}$. An error $\delta$ is also introduced such that
        \begin{equation}
            \delta = 20 \log \left(\frac{\lVert \boldsymbol{E}_{s, \mathrm{model}} - \boldsymbol{E}_{s, \mathrm{ref}} \rVert}{\max{\big(\lVert\boldsymbol{E}_{s, \mathrm{ref}} \rVert \big)}}\right)
        \end{equation}
        where $\boldsymbol{E}_{s, \mathrm{model}}$ designates the field~(\ref{equ_total_field}) and $\boldsymbol{E}_{s, \mathrm{ref}}$ the field calculated with the reference MLFMM simulation.

        \subsection{Thick finite cylinder}
            The model is first tested with a simple case, a cylinder of length $L = 20\lambda$ and radius $R = 3\lambda$ which is represented in Figure~\ref{fig_thick_cylinder}. Both MLFMM and large elements Physical Optics (PO) solvers are tested. The objective is to show that our model outperforms PO in terms of accuracy, for a marginal increase of the computation time. The study of this simple case also allows to first introduce the calculated quantities as well as the graphical representations that will be used throughout the rest of the article on more complex cases.

            The forward scattering (the scattered field in the positive $\hat{\boldsymbol{x}}$ direction) calculated with the model in the $x = 2d_\mathrm{far}$ plane is represented in Figure~\ref{fig_thick_cylinder} (b) and the relative difference $\delta$ between model and simulation in Figure~\ref{fig_thick_cylinder} (c). This representation of the fields shows an overall excellent agreement between model and reference simulation as the maximum value of $\delta$ is -23~dB. The differences are expected to be mostly caused by the top and bottom surfaces and edge effects of the cylinder in the simulation which are absent in the model. It has been observed that the thinner the cylinders, the less significant these differences are.
            
            A cut in the spherical far field at $\phi = 0^\circ$ as well as an  additional cut at $z = 0$ in the $x = 2d_\mathrm{far}$ plane are represented in Figure~\ref{fig_thick_cylinder} (d) and (e) respectively. The PO result is added to the comparisons. The model behaves significantly better than PO in both representations and strongly matches the MLFMM reference simulation in the $z = 0$ cutplane while their solution times are both very short as they are less than one thousandth of a second. For this reason, PO results are ommited in the remaining field comparisons.

        \subsection{Set of regularly arranged cylinders}
            One step further towards generality, sets of 9 regularly arranged cylinders are studied. The very same calculations, plots and comparisons as for the thick cylinder case are performed. In a first subcase, cylinders are of length $L = 40\lambda$, of radius $R = \lambda$ and are $20\lambda$ away from another. The scene and results are shown in Figure~\ref{fig_9_cylinders}.
            
            The forward scattering in the $x = 2d_\mathrm{far}$ cutplane clearly shows the lattice effects with the apparition of characteristic grating lobes $\hat{\boldsymbol{y}}$. Again, few differences are visible between model and reference simulation as $\delta$ maximum value is also -23~dB. The $x$ and $\phi$ cuts show a very good agreement between model and reference simulations, especially in aeras where the scattered field is strong.
            
            In a second subcase, the model is also tested on the same array of cylinders but with a random cylinder radii distribution with $R_p \in [0.5\lambda, 3\lambda]$. For the sake of brevity, simulation results of that subcase are not extensively shown here. Nevertheless, conclusions are similar to those of the previous subcase as it is observed that the lattice structure still influences greatly the radiation pattern in such a case, the strong scattered field amplitudes are still very well predicted and the greatest error $\delta$ is of -18~dB.
          
    \subsection{Set of randomly placed cylinders}
        A configuration, whose results are also briefly summarized here, where 9 identical cylinders of individual length $L = 40\lambda$ and radius $R = \lambda$, randomly arranged in a $100\lambda \times 100\lambda$ square is tested. In such a case, most of the scattered power is along the incident plane-wave direction with a clear maximum along $\hat{\boldsymbol{x}}$. The maximum value of $\delta$ for this case is -17~dB.
        
        Increasing in complexity, a more general case is finally explored: the set of randomly arranged cylinders of random radii. The situation is identical from the previous one but cylinders radii are now those of Table~\ref{tab_randc_randR}.
        \begin{table}
            \centering
            \caption{Radii of the random radii randomly placed cylinders.}\label{tab_randc_randR}
            \begin{tabular}{cc|cc|cc}
                \toprule
                $p$ & $R_p$ ($\lambda$) & $p$ & $R_p$ ($\lambda$) & $p$ & $R_p$ ($\lambda$) \\
                \midrule
                $1$ & $2.858$ & $4$ & $2.405$ & $7$ & $2.991$ \\
                $2$ & $2.480$ & $5$ & $0.800$ & $8$ & $1.561$ \\
                $3$ & $1.130$ & $6$ & $1.129$ & $9$ & $1.474$ \\
                \bottomrule
            \end{tabular}
        \end{table}
        The results for this configuration are presented in Figure~\ref{fig_9_cylinders_randc_randR}.
        
        A clear maximum of scattering is observed along the incident plane wave direction with a rapid decay around it. The reference to model agreement remains very good, especially in the spherical cut, although fast erratic variations are observed in the radiation patterns. The greatest value of the error $\delta$ is -15~dB for this case. Like in any situation where several cylinders are involved, the error maximal values are located in the reference plane.
        
\begin{figure}
            \centering
            \includegraphics{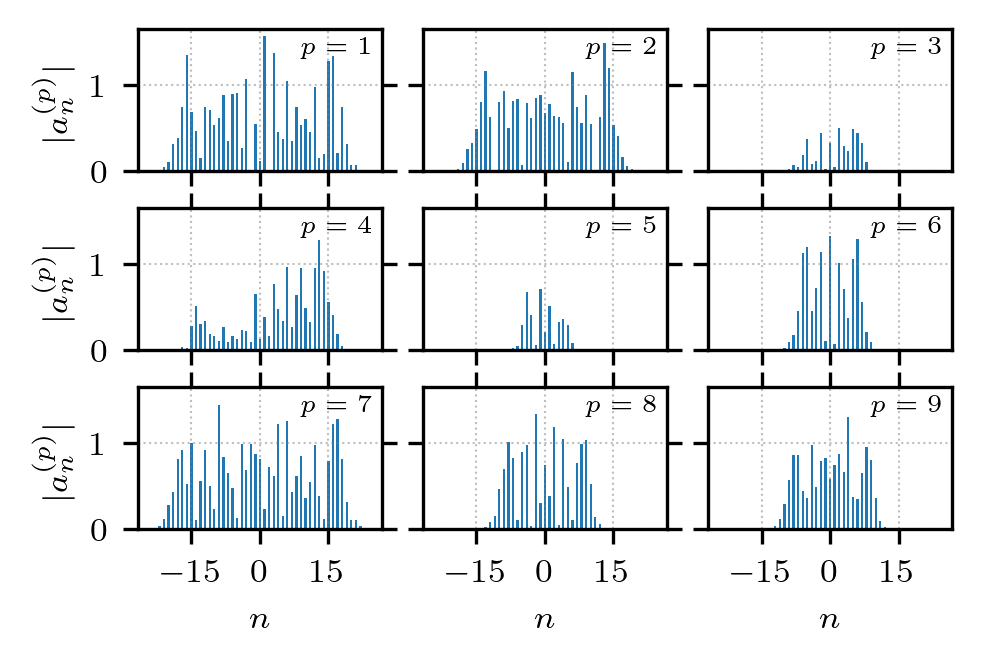}
            \caption{Harmonic coefficients repartition in the case of the set of 9 randomly placed cylinders of random radii.}\label{fig_9_cylinders_randc_randR_coefs}
        \end{figure}

        For this very last case study, the $a_n^{(p)}$ coefficient masses are represented in Figure~\ref{fig_9_cylinders_randc_randR_coefs}.
        This allows to quickly discuss the scattering contributions of the scene cylinders.
        \begin{itemize}
            \item As seen in the scene representation, cylinders 1, 2 and 7 are of larger radius than other cylinders so they logically show a greater total number of non-zero harmonics.
            \item Cylinders 1 and 7 harmonic modules are greater than cylinder 2 modules as they stand in front of other cylinders of the configuration.
            \item Cylinders 3 and 5, despite having similar geometrical characteristics and being at roughly the same $x$ position than cylinders 6 and 9 show much lower harmonic modules. They are shadowed by front cylinders 1 and 7 that are at similar $y$ positions.
        \end{itemize}

    \section{Computational times}
        In this final section, computational times from the reference solution and the proposed model solution are compared through a parametric study on cylinders radii, since this quantity is expected to drive the sizes of the systems to solve.

        \subsection{Model complexity}
        Solution times are splited into two parts: the currents computation time $t_j$ and the fields computation time $t_e$. In our model, the $t_j$ duration is mostly associated with solving the linear system and is then primarily influenced by the size of its matrix. The $t_e$ duration is associated with the current radiation at given observation points and is then primarily influenced by the number of points in the observation grid. In this article, all values of $t_e$ are given for 1000 points.

        Practically, for the model calculations, $t_e$ is heavily influenced by the calculation of $J_n\big(kR_p\sin(\theta_p)\big)$ in the summation term~(\ref{equ_summation_term}). As it can be seen from this same equation the model requires the evaluation of ${\mathscr{N}_p = 2N^{(p)}+1}$ Bessel and Hankel functions for each unique value of $R_p$ in the scattering scene and, additionnaly, of $\mathscr{N}_p \times n_\theta$ Bessel functions for a $\theta$ grid of $n_\theta$ points. This last restriction, on the spherical grid of points, is by far the most expensive part of the calculation. The current computation time $t_j$ corresponds to the inversion of $\boldsymbol{M}$ in equation~(\ref{equ_linear_system}) which is of size ${\left(\sum_p \mathscr{N}_p\right)}^2$.

        \subsection{Parametric study}
            The configuration of Figure~\ref{fig_9_cylinders}~(a) is chosen to perform a parametric study on the shared radius $R$ of the cylinders. Simulations are performed on a 3.90~GHz Intel Xeon W-2245 CPU along with 256~Go RAM resources. For both reference and model, problems are solved using no parallelisations as the code implementing the proposed model do not include any computation parallelisation feature yet for this preliminary study. Cylinders are simulated for $R$ values up to $5\lambda$. 
            
        \subsection{Results}
            Above the $R = 5\lambda$ value, the computer used does not provide sufficient RAM resources to solve the problem with the reference MLFMM computation, where it does with the proposed model computation. This limitation itself is already a result of the parametric study as it shows one clear advantage of the model computation.

\begin{figure}
                \centering
                \includegraphics{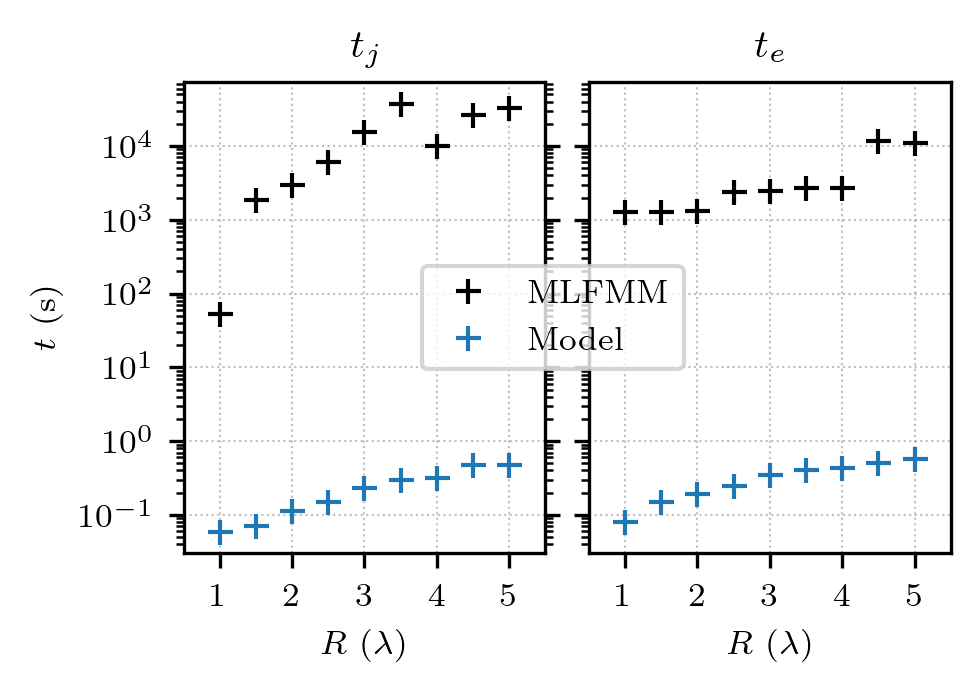}
                \caption{Computational times comparisons between reference and model solutions. Necessary CPU time for the computation of (left) currents and (right) fields at 1000 points of observation.}\label{fig_R_parametric}
            \end{figure}

            Computational times in seconds are gathered in Figure~\ref{fig_R_parametric}. For both currents and fields computation times, the proposed model is faster than the reference solution by 5 orders of magnitude. For this specific parametric study, no model computation time exceeds a second whereas the average reference solution time is around few hours. The field computation times $t_e$ follow similar variations in both cases but the current computation times $t_j$ increases more rapidly in the reference solution case.
        
    \section{Conclusion}
        In this publication, a closed-form solution for scattering problems involving arbitrarily placed co-oriented cylinders of arbitrary radii and arbitrary but large lengths has been developed. Various configurations have been studied and the model has shown very good agreement with the reference MLFMM solutions, the relative error never exceedeing -15~dB. Scattered field variations are well predicted and very accurate for the strongest field amplitudes. The model is faster than reference solution by 5 orders of magnitude. The bottleneck of the model computational time is on the number of point in the desired spherical $\theta$ grid.
                
        We have restricted this study to metallic cylinders for the sake of simplicity, but the method remains the same for dielectrics as long as the suitable current density is taken such as the one calculated in~\cite{Barber1975}, for example.

        The far field Stratton-Chu formulation has been used for each cylinder. This means that the calculation is valid only in far field region of each cylinder. Nevertheless, it is not necessary to be in the far field zone of the set of cylinders. In cases where the far field region of one cylinder is still too far from an application point of view, the cylinder can be subdivided into smaller cylinders whose heights are fractions of $L_p$ and the integral evaluated on these subdomains.

    % \section*{Data availability}
    %     The repository~\cite{Douvenot2022} contains routines that allow one to generates the exact same reference simulations (by the means of \texttt{.lua} files readable by Feko) in order to generate the very same simulation data. Simulation data can still be provided to the reader upon reasonable request.

    \section*{Acknowledgments}
        This work has been founded by DGAC-DSNA-DTI.\@

    \section*{CRediT authorship contribution statement}
        \textbf{Matthieu Elineau} --- Conceptualization, Formal Analysis, Investigation, Methodology, Software, Validation, Visualization, Writing - Original Draft.

        \textbf{Lucille Kuhler} --- Conceptualization, Formal Analysis, Investigation, Methodology, Software, Validation, Writing - Review \& Editing.

        \textbf{Alexandre Chabory} --- Conceptualization, Formal Analysis, Funding Acquisition, Investigation, Methodology, Project Administration, Software, Supervision, Validation, Writing - Review \& Editing.

    \setcounter{section}{0}
    \renewcommand\thesection{\Alph{section}}
    \numberwithin{equation}{section}

    \section{Radiation integral calculation}\label{supp_radiation_integral}
    This appendix details the developments involved in the calculation of the radiation integral~(\ref{equ_interm_es}) in the article main body. We call $I$ the integral to be solved within the summation of the radiation integral~(\ref{equ_interm_es}). It can be expressed as the product of two independent integrals with respect to $z_p'$ and $\phi_p'$ such that
    \begin{equation}
        \begin{split}
            I &= I_{z_p'} \times I_{\phi_p'}\\
            &= \int_{z_p' = h_p}^{z_p' = h_p+L_p} e^{jk\cos(\theta_p)z_p'}\mathrm{d}z_p'\\
            &\quad\times\int_{\phi_p' = 0}^{\phi_p' = 2\pi} e^{jn\phi_p'} e^{jk\sin(\theta_p) R_p\cos(\phi_p' - \phi_p)}\mathrm{d}\phi_p'.
        \end{split}
    \end{equation} 

\subsection{Integral with respect to $z_p'$}
    The calculation of $I_{z_p'}$ is straightforward and gives
    \begin{equation}\label{equ_I_{z_p'}}
        I_{z_p'} = L_p \frac{\sin(Z_p)}{Z_p}e^{jk\cos(\theta_p)\big(h_p+\frac{L_p}{2}\big)}.
    \end{equation}
    with $Z_p = \frac{k\cos(\theta_p)L_p}{2}$.

\subsection{Integral with respect to $\phi_p'$}
    The second part of the product is
    \begin{equation}
        I_{\phi_p'} = \int_{\phi_p' = 0}^{\phi_p' = 2\pi} e^{jn\phi_p'} e^{jk R_p\sin(\theta_p)\cos(\phi_p' - \phi_p)}\mathrm{d}\phi_p'.
    \end{equation}
    With the substitution $\Phi = \phi_p' - \phi_p$, its expression becomes
    \begin{equation}
        I_{\phi_p'} = \int_{\Phi = -\phi_p}^{\Phi = 2\pi - \phi_p} e^{jn(\Phi + \phi_p)} e^{jk R_p\sin(\theta_p)\cos(\Phi)}\mathrm{d}\Phi.
    \end{equation}
    The integral is for $2\pi$ periodic functions over one full rotation on the circle. It allows to shift the two integration bounds by a same quantity $+\phi_p$ such that
    \begin{equation}\label{equ_interm_i_phi}
        \begin{split}
            I_{\phi_p'} &= \int_{\Phi = 0}^{\Phi = 2\pi} e^{jn(\Phi + \phi_p)} e^{jk R_p\sin(\theta_p)\cos(\Phi)}\mathrm{d}\Phi\\
            &= e^{jn\phi_p} \int_{\Phi = 0}^{\Phi = 2\pi} e^{jn\Phi} e^{jk R_p\sin(\theta_p)\cos(\Phi)}\mathrm{d}\Phi.
        \end{split}
    \end{equation}
    The remaining integral is close to the well know form~(5--102) in~\cite{Harrington2001}
    \begin{equation}
        J_n(z) = \frac{j^n}{2\pi}\int_{0}^{2\pi}e^{-jn\Phi}e^{-jz\cos(\Phi)}\mathrm{d}\Phi
    \end{equation}
    which can alternatively be written, with ${z = -k R_p \sin(\theta_p)}$ and for $J_{-n}$,
    \begin{equation}
        \begin{split}
            J_{-n}\big(-k R_p \sin(\theta_p)\big) &= \frac{j^{-n}}{2\pi}\\
            &\quad\times\int_{0}^{2\pi}e^{jn\Phi}e^{jk R_p \sin(\theta_p)\cos(\Phi)}\mathrm{d}\Phi.
        \end{split}
    \end{equation}
    Using the identity of Bessel functions $J_{-n}(z) = J_n(-z)$ and isolating the integral in the right hand side of the equation we obtain
    \begin{equation}
        \begin{split}
            \ J_n\big(-k R_p \sin(\theta_p)\big)&\frac{2\pi}{j^{-n}}
            = \ 2\pi j^n J_n\big(-k R_p \sin(\theta_p)\big)\\
            = &\int_{0}^{2\pi}e^{jn\Phi}e^{jk R_p \sin(\theta_p)\cos(\Phi)}\mathrm{d}\Phi
        \end{split}
    \end{equation}
    which is precisely the integral to be calculated in equation~(\ref{equ_interm_i_phi}). Inserting this expression in $I_{\phi_p'}$ we obtain
    \begin{equation}\label{equ_I_phi}
        I_{\phi_p'} = 2\pi j^n J_n\big(k R_p\sin(\theta_p)\big)e^{jn\phi_p}.
    \end{equation}

\subsection{Final expression of the scattered field}
    The integral $I$ is obtained with the product of the integrals~(\ref{equ_I_{z_p'}}) and~(\ref{equ_I_phi}) which is
    \begin{equation}
        \begin{split}
            I &= 2L_p\pi\frac{\sin(Z_p)}{Z_p}j^n J_n\big(k R_p\sin(\theta_p)\big)e^{jn\phi_p}\\
            &\quad\times e^{jk\cos(\theta_p)\big(h_p+\frac{L_p}{2}\big)}.
        \end{split}
    \end{equation}
    This integral is inserted into the scattered field expression~(\ref{equ_I_{z_p'}}) leading to
    \begin{equation}
        \begin{split}
        E_{s, \theta_p}^{(p)} &= \frac{k R_p L_p}{4} \frac{\sin(Z_p)}{Z_p}\frac{e^{-jkr_p}}{r_p}e^{jk\cos(\theta_p)\big(h_p+\frac{L_p}{2}\big)}\\
        &\quad\times\sum_{n}\Big\{a_n^{(p)}H_n^{(2)}(kR_p)\left[\mathscr{H}_n(kR_p) - \mathscr{J}_n(kR_p)\right]\\
        &\qquad\qquad\quad \times j^n J_n\big(k R_p\sin(\theta_p)\big)e^{jn\phi_p}\Big\}
        \end{split}
    \end{equation}
    which is the final scattered field expression~(\ref{equ_e_theta_p}) after identification of $\mathscr{E}^{(p)}$ as the multiplying factor of the sum and $\mathscr{S}_n^{(p)}$ as the summation term.

\printbibliography%
\end{document}